\documentclass[preprint,preprintnumbers,amsmath,amssymb]{revtex4}

\usepackage{graphicx}
\usepackage{bm}
\usepackage{amssymb}

\begin{document}

\title{Longitudinal Ion Acceleration from High-Intensity Laser Interactions with Underdense Plasma}

\author{L.~Willingale$^1$}
\author{S.~P.~D. Mangles$^1$}
\author{P.~M.~Nilson$^{1 \dag}$}
\author{R.~J.~Clarke$^2$}
\author{A.~E.~Dangor$^1$}
\author{M.~C.~Kaluza$^{1 \ddag}$}
\author{S.~Karsch$^3$}
\author{K.~L.~Lancaster$^2$}
\author{W.~B.~Mori$^5$}
\author{J.~Schreiber$^{3,4}$}
\author{A.~G.~R. Thomas$^1$}
\author{M.~S.~Wei$^1$}
\author{K.~Krushelnick$^{1 \ast}$}
\author{Z.~Najmudin$^1$}

\affiliation{$^1$ Blackett Laboratory, Imperial College London, London SW7 2AZ, United Kingdom}
\affiliation{$^2$ Central Laser Facility, Rutherford Appleton Laboratory, Chilton, Oxon, United Kingdom}
\affiliation{$^3$ Max-Planck Institut f\"ur Quantenoptik, Garching, Germany}
\affiliation{$^4$ Ludwig-Maximilians-Universit\"at M\"unchen, Am Coulombwall 1, D-85748 Garching, Germany}
\affiliation{$^5$ Department of Physics \& Astronomy \& Department of Electrical Engineering, UCLA, Los Angeles CA, USA}

\begin{abstract}
Longitudinal ion acceleration from high-intensity ($I \sim 10^{20} \; \rm{Wcm}^{-2}$) laser interactions with helium gas jet targets ($n_{e} \approx 0.04 n_{c}$) have been observed.
The ion beam has a maximum energy for $\rm{He}^{2+}$ of $(40^{+3}_{-8}) \; \rm{MeV}$ and was directional along the laser propagation path, with the highest energy ions being collimated to a cone of less than $10^{\circ}$.
2D particle-in-cell simulations have been used to investigate the acceleration mechanism.
The time varying magnetic field associated with the fast electron current provides a contribution to the accelerating electric field as well as providing a collimating field for the ions.
A strong correlation between the plasma density and the ion acceleration was found.
A short plasma scale-length at the vacuum interface was observed to be beneficial for the maximum ion energies, but the collimation appears to be improved with longer scale-lengths due to enhanced magnetic fields in the ramp acceleration region.
\end{abstract}

\maketitle

\section{Introduction}

Laser intensities currently available can accelerate electrons directly with the laser field to energies of many times their rest mass, $m_{e} = 511 \; \rm{keV}$, making them highly relativistic.
The comparatively heavy ions only become relativistic as a direct consequence of the laser field when the normalised vector potential, $a_{0}$, exceeds $1836$, which corresponds to $I \lambda^{2} = 4.6 \times 10^{24} \; \rm{Wcm}^{-2} \mu \rm{m}^{2}$.
Therefore experimental investigation is currently confined to non-relativistic intensities for ions and their movement is governed by the bulk movement of electrons.
However, extremely large accelerating gradients ($> 10^{12} \; \rm{Vm}^{-1}$) can be supported in plasmas, which means that the scale of such plasma based accelerators is reduced by orders of magnitude compared with conventional designs and so ions can be readily accelerated to non-relativistic energies with current laser systems.

The non-relativistic critical plasma density for a laser with a frequency $\omega_{L}$ is $n_{c} = \epsilon_{0} m_{e} \omega_{L}^{2} / e^{2}$ and is the plasma density above which the laser is unable to propagate.
Above this density the plasma is known as overdense and below is underdense.
In high-intensity laser interactions with solid targets (usually overdense), electrons can be accelerated through a number of mechanisms, including vacuum heating \cite{Brunel_PRL_1987}, $\mathbf{j} \times \mathbf{B}$ heating \cite{Kruer_PoF_1985} and direct laser acceleration \cite{Mangles_PRL_2005}.
Relativistic electrons can be generated in this way and as they attempt to leave the target and move into the surrounding vacuum a large space-charge electric field is generated.
Ions can be accelerated to $10$s of MeV in a direction normal to the target surface.
Therefore this is known as the target normal sheath acceleration (TNSA) mechanism.
Solid targets are used in the majority of proton and ion acceleration experiments, taking advantage of the thin ($\sim \rm{nm}$) hydrocarbon layers formed on the surface of targets as the source of protons \cite{Clark_PRL_2000, Snavely_PRL_2000}.
The production of high-flux, small-emittance \cite{Cowan_PRL_2004} and perhaps even quasi-mono-energetic \cite{Schwoerer_Nature_2006, Hegelich_Nature_2006} proton beams from laser-plasma interactions have numerous and diverse potential applications; less invasive radiotherapy \cite{Bulanov_PLA_2002}, isotope production for positron emission tomography (PET) \cite{Spencer_NIMPR_2001}, proton radiography of electromagnetic fields \cite{Borghesi_LPB_2002}, high brightness neutron sources or to deliver the ignition energy in the fast ignitor inertial confinement fusion concept \cite{Roth_PRL_2001}.

Underdense plasmas have provided an efficient medium for the acceleration of electrons to high energies: direct laser acceleration has produced electrons with energies of up to $350 \; \rm{MeV}$ with temperatures exceeding the ponderomotive potential \cite{Mangles_PRL_2005} and laser wakefield acceleration has demonstrated that monoenergetic electron bunches of $\sim 1 \; \rm{GeV}$ can be generated \cite{Leemans_NP_2006}.
Ion acceleration from these underdense plasmas has previously been limited to the transverse direction with ion acceleration via the ponderomotive Coulomb explosion \cite{Krushelnick_PRL_1999} and shock acceleration \cite{Wei_PRL_2004} mechanisms.
More recently a beam of multi-MeV ions has been measured from an underdense plasma with an ultraintense laser pulse \cite{Willingale_PRL_2006, Willingale_PRLcomment_2007} and this paper provides further insight into the interaction and ion acceleration mechanism.
The experimental results presented in reference \cite{Willingale_PRL_2006} are reviewed and followed by particle-in-cell (PIC) simulation results investigating the acceleration mechanism and the effect of the plasma density and scale-length on the ion acceleration.

\section{Experiment}

\subsection{Experimental set up}
The experiment was performed using the Vulcan Petawatt laser \cite{Vulcan_PW} at the Rutherford Appleton Laboratory.
The laser pulse had a duration of $\tau_{L} \approx 1.0 \; \rm{ps}$ full-width-half-maximum (FWHM) with energy of up to $340 \; \rm{J}$ on target to give a maximum power of $340 \; \rm{TW}$.
The central wavelength of the Vulcan Petawatt laser is $\lambda_{0} = 1.054 \; \mu \rm{m}$ so the non-relativistic critical density is $n_{c} = 1.0 \times 10^{21} \; \rm{cm}^{-3}$.
An $f/3$ off-axis parabolic mirror focused the laser to spot with a FWHM diameter of $7 \; \mu \rm{m}$ to provide a cycle averaged peak vacuum intensity of $1.5 \times 10^{20} \; \rm{Wcm}^{-2}$, which corresponds to a peak $a_{0} \approx 21$.
The contrast ratio was $\sim 10^{-5}$.

The laser was focused to the edge of the gas flow from a $2 \; \rm{mm}$ diameter supersonic nozzle.
Supersonic nozzles were used to ensure a uniform density profile over a given distance and a reasonably sharp density ramp ($< 250 \; \mu \rm{m}$) both at the front and back of the gas jet.
The density profile of the gas jet was determined prior to the high-intensity shots by interferometry.
The backing pressure could be varied so that the electron density of the fully ionised plasma could be set between $(0.7$--$4.0) \times 10^{19} \; \rm{cm}^{-3}$.
The frequency separation of the forward Raman scattered laser spectra ($\triangle \omega = \omega_{pe} = \sqrt{n_{e} e^{2} / m_{e} \epsilon_{0}}$) confirmed the electron density $n_{e}$ of the interaction.

Ion energy spectra were taken at four different angles from the laser axis to measure the angular emission.
Thomson ion spectrometers were placed at $90^{\circ}$ and $45^{\circ}$ to the laser propagation direction.
In addition, there were two charged particle spectrometers at $10^{\circ}$ and $0^{\circ}$ to the laser propagation direction.
Though primarily to measure the electron spectra, the open geometry of the magnetic spectrometers also allowed the measurement of ions and other positively charged particles.
Ions are deflected in the opposite direction from electrons in the magnetic field.
The deflection of ions in a magnetic field depends simply on their mass-to-charge ratio and their momentum, thus allowing their energy to be determined.
The nuclear track detector CR39, which is insensitive to positrons, was used to detect the ions.

For a helium gas target, there are two possible ion species, $\rm{He}^{1+}$ and $\rm{He}^{2+}$.
In the Thomson spectrometer, these different charge-to-mass ratio species are separated spatially by an electric field in addition to the magnetic field.
In the magnetic spectrometers, the different species are not separated; therefore, it was necessary to distinguish between the species by the pit size for all ions deflected by a similar amount.
At a particular point on the detector, both ion species may be present but the $\rm{He}^{2+}$ will have more energy.
The higher the energy of an ion, the further into the CR39 it will travel before it is stopped and therefore the damage will be deeper into the material.
The CR39 is etched in a NaOH solution so that pits are formed in the damaged regions.
Smaller pits on the surface are the result of deeper damage.
Hence at a particular point, using the electron spectrometers, the smaller pits will be from the $\rm{He}^{2+}$ and the larger ones from $\rm{He}^{1+}$ ions, allowing the species to be differentiated.
Electron spectra were measured simultaneously during the shot using image plate detectors as reported previously \cite{Mangles_PRL_2005}.

\subsection{Experimental results}

For plasma densities of $\lesssim 2 \times 10^{19} \; \rm{cm}^{-3}$, energetic ions were found to be emitted primarily in the transverse direction consistent with previous measurements \cite{Krushelnick_PRL_1999, Wei_PRL_2004}.
However, above this density, a clear, reproducible signal of ions was observed in the $0^{\circ}$ and $10^{\circ}$ spectrometers.
Presented here is a shot which had an on-target energy of $340 \; \rm{J}$ (peak $I_{0} \approx 6 \times 10^{20} \; \rm{Wcm}^{-2}$) and was incident on a helium plasma with $n_{e} \approx 4 \times 10^{19} \; \rm{cm}^{-3}$.

\begin{figure}
\begin{center}
\includegraphics[width=5cm]{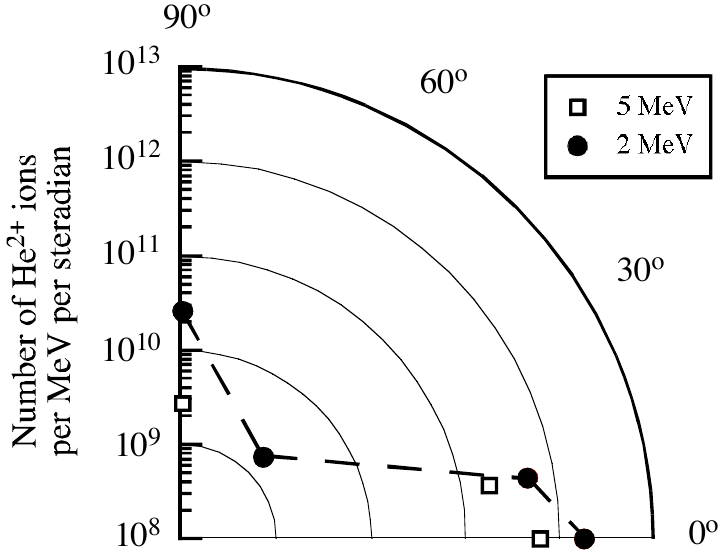}
\caption{A polar plot showing the number of $\rm{He}^{2+}$ ions at different diagnostic angles for $2 \; \rm{MeV}$ and $5 \; \rm{MeV}$ ions.}
\label{figure_polar}
\end{center}
\end{figure}

On this shot, the on-axis ($0^{\circ}$) electron spectrum extended to $66 \pm 2 \; \rm{MeV}$, with a characteristic temperature of $T_{e} \approx 7.4 \; \rm{MeV}$.
The $10^{\circ}$ electron spectrum was a little hotter, $T_{e} \approx 9.8 \; \rm{MeV}$, possibly due to filamentation and hosing in the interaction.
In the transverse direction, the $\rm{He}^{2+}$ had a maximum energy of $(7.8 \pm 0.6) \; \rm{MeV}$ and  the $\rm{He}^{1+}$ ions had a maximum energy of $(3.4 \pm 0.3) \; \rm{MeV}$.
In the longitudinal direction, the $\rm{He}^{2+}$ had a maximum energy of $(40^{+3}_{-8}) \; \rm{MeV}$ and  the $\rm{He}^{1+}$ ions had a maximum energy of $(10^{+3}_{-2}) \; \rm{MeV}$.
Fig.\ \ref{figure_polar} shows a polar plot of the number of $\rm{He}^{2+}$ ions with $2 \; \rm{MeV}$ and $5 \; \rm{MeV}$ at the different diagnostic angles.
The $\rm{He}^{2+}$ ions have formed two lobes, one in the forward direction along the laser axis and the other in the radial direction.
Hence, it is deduced that there is a well collimated ion beam in the forward direction with a divergence angle of less that $10^{\circ}$.

\section{Simulations}

The 2D3V particle-in-cell code Osiris \cite{osiris} was used to investigate the ion acceleration mechanism and the effect of various parameters.
The simulations were performed on the 48 node `Caesar' cluster and on the CX1 supercomputer at Imperial College.
Stationary simulation boxes were used to observe the plasma even after the laser has passed.
The stationary simulation box was $251 \times 251 \; \mu \rm{m}$, with a resolution of $20.9$ cells /$\lambda$ in the longitudinal ($x$) direction and $12.6$ cells /$\lambda$ in the transverse direction ($y$).
Since ionisation effects are not included in this version of the code, a $\rm{He}^{2+}$ plasma was used to avoid complications due to balance of the different charge states.
The ion acceleration was investigated with various values of plasma electron density, of $0.01\; n_{c}$, $0.05 \; n_{c}$, $0.1\; n_{c}$, $0.3\; n_{c}$, $0.5\; n_{c}$, $0.75\; n_{c}$ and $0.9\; n_{c}$ where $n_{c}$ is the non-relativistic critical density.
The density profile for these simulations had a $71\; \mu \rm{m}$ long linear density ramp at the front of the plasma, $25\; \mu \rm{m}$ of plasma at maximum density and a $71\; \mu \rm{m}$ density ramp at the back of the target.
The vacuum then extended a further $84\; \mu \rm{m}$ behind the plasma.
The laser pulse was linearly polarised with the laser's electric field in the $y$-direction, with a full-width-half-maximum pulse length of $\tau_l = 500\; \rm{fs}$ and a wavelength $\lambda_{0} = 1.053\; \mu \rm{m}$.
It was focused to a full-width-half-maximum diameter spot $\rm{w}_{0} = 8\; \mu \rm{m}$ at the top of the front density ramp, to give a peak normalised vector potential $a_{0} \approx 15$.

Another parameter investigated was the length of the density ramp at the rear of the target.
In solid target ion acceleration, density ramps formed by pre-plasma at the rear of the target have been found to be detrimental to ion acceleration \cite{Kaluza_PRL_2004}.
The reason for this is, if the pre-plasma has a scale length of greater than the Debye length, $\lambda_{D} = \sqrt{\epsilon_{0} k_{B} T / e^{2} n_{e0}}$, the electrons are able to move to shield ions from the accelerating electric field preventing efficient acceleration.
For solid targets $\lambda_{D}$ will be of the order of nanometers, but for the underdense targets discussed here, $\lambda_{D}$ is of the order of micrometers.
Taking the electron temperature to be $T_{e} = 7 \; \rm{MeV}$ and the electron density to be $4 \times 10^{19} \; \rm{cm}^{-3}$ gives a $\lambda_{D} \sim 3 \; \mu \rm{m}$.
It is therefore expected that the density ramp on the gas jet targets used in the experiment will have a detrimental effect on the ion acceleration, and that the simulations can provide an approximate scaling with the ramp length.
For these simulations, the simulation box was $598 \times 168 \; \mu \rm{m}$ with a resolution of $19.8$ cells /$\lambda$ in the longitudinal ($x$) direction and $9.4$ cells /$\lambda$ in the transverse direction ($y$).
Again, the ion species simulated was $\rm{He}^{2+}$ and the initial plasma electron density used was $0.05 n_{c}$ (equivalent to $5 \times 10^{19} \; \rm{cm}^{-3}$, which is similar to the experiment).
The lengths of the density ramps at the rear vacuum plasma interface were $0 \; \mu \rm{m}$, $100 \; \mu \rm{m}$ and $200 \; \mu \rm{m}$.
The centre of each of these ramps was at the same position so that the total amount of plasma the laser travels through is the same for all of the simulations.
The laser pulse was linearly polarised with the laser's electric field in the $y$-direction, with $\tau_{L} = 500\; \rm{fs}$ and $\lambda_{0} = 1.053\; \mu \rm{m}$.
It was focused to a $\rm{w}_{0} = 2.4\; \mu \rm{m}$ at the top of the front density ramp to give an $a_{0} \approx 15$.

The simulations assume an initially fully-ionised plasma and, in the case of an interaction with a relativistic laser pulse, the effects should be minimal as the front of the laser pulse will rapidly ionise any atoms it passes by.
Care must be taken, when examining the simulation results when periodic boundary conditions have been used, to make sure the effect of any recirculation of particles is minimal, or to look at the result before recirculation occurs.
The boundary conditions for these simulations were periodic in the transverse direction.

\subsection{Ion acceleration and collimation mechanism}

\begin{figure}
\begin{center}
\includegraphics[width=8cm]{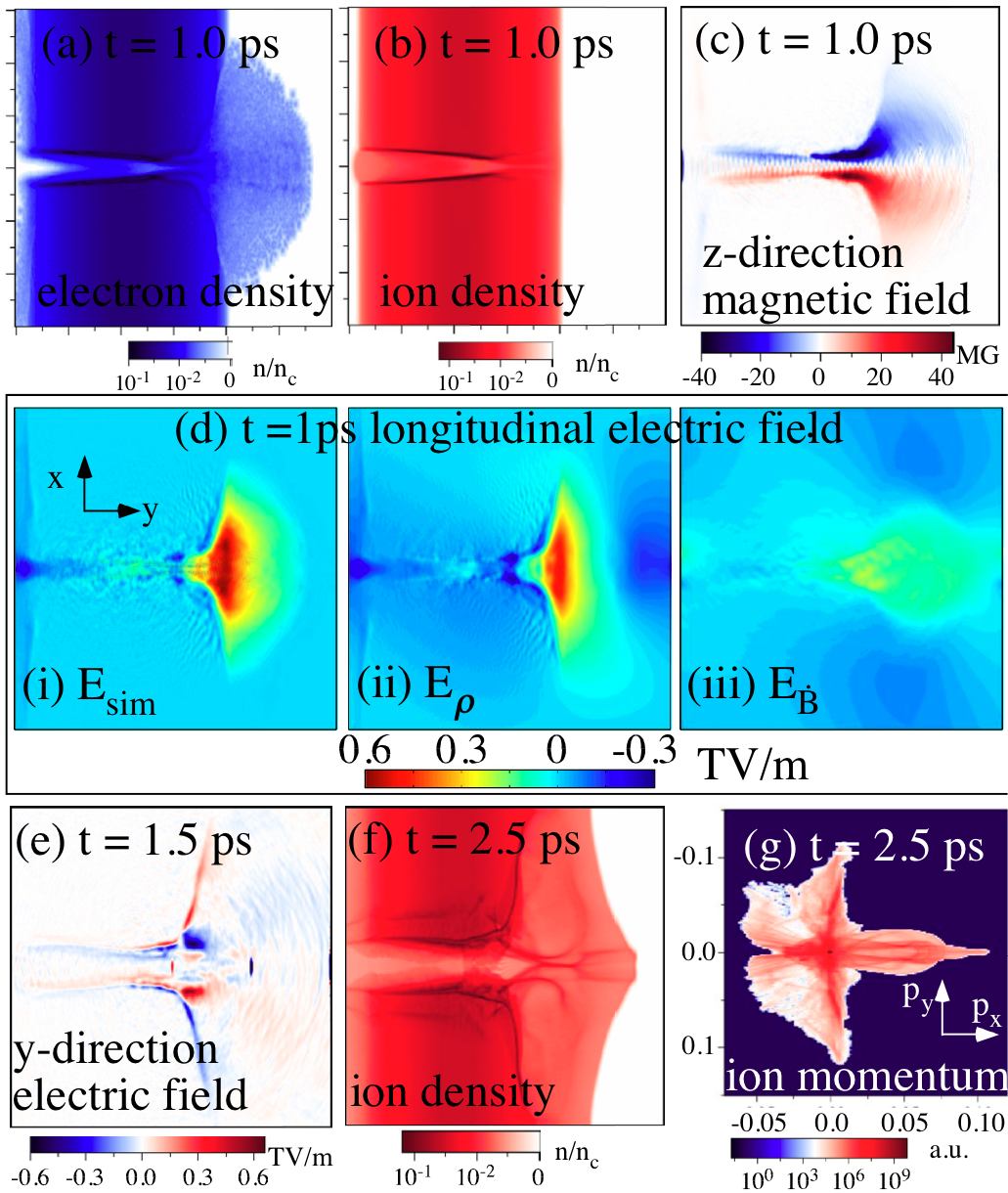}
\caption{(a) The electron density, (b) the ion density and (c) the azimuthal magnetic field at a time of $1 \; \rm{ps}$ into the simulation. (d) The longitudinal electric field components at a time of $1 \; \rm{ps}$ into the simulation: (i) the total longitudinal electric field seen in the simulation, $E_{sim}$, (ii) the charge separation contribution, $E_{\rho}$ and (iii) the time varying magnetic field contribution, $E_{\dot{B}}$. (e) The radial electric field at $1.5 \; \rm{ps}$, (f) the ion collimation at $2.5 \; \rm{ps}$ and (g) shows the $\rm{He}^{2+}$ ion $p_{x}$ against $p_{y}$ momentum at $2.5 \; \rm{ps}$.}
\label{osiris_accn_figure}
\end{center}
\end{figure}

After $1 \; \rm{ps}$ into the simulation, the laser pulse starts to emerge from the rear of the plasma into vacuum.
The accelerated electrons also move out into the vacuum region at the rear of the plasma.
The ions do not immediately respond to the movement of the electrons due to their large mass and therefore a large space charge electric field in set up.
The electron and ion densities can been seen in fig.\ \ref{osiris_accn_figure} (a) and (b) at a time of $1 \; \rm{ps}$ into the simulation, which shows the charge separation.
The longitudinal electric field in the simulation, $E_{\rm{sim}}$ is at its maximum value of $\approx 0.7 \; \rm{TV/m}$ at $1 \; \rm{ps}$ and is shown in fig.\ \ref{osiris_accn_figure} (d) (i).
The simulation ion and electron data can be used to calculate the charge density and therefore the longitudinal electric field due to charge separation, $E_{\rho}$, can be found.
The charge separation contribution to the longitudinal electric field at $1 \; \rm{ps}$ into the simulation is shown in fig.\ \ref{osiris_accn_figure} (d) (ii) and has a maximum value of $\approx 0.5 \; \rm{TV/m}$.
Although the charge separation makes up the largest contribution to the longitudinal electric field, there is a discrepancy so there must be an additional electric field generation mechanism.

\begin{figure}
\begin{center}
\includegraphics[width=8cm]{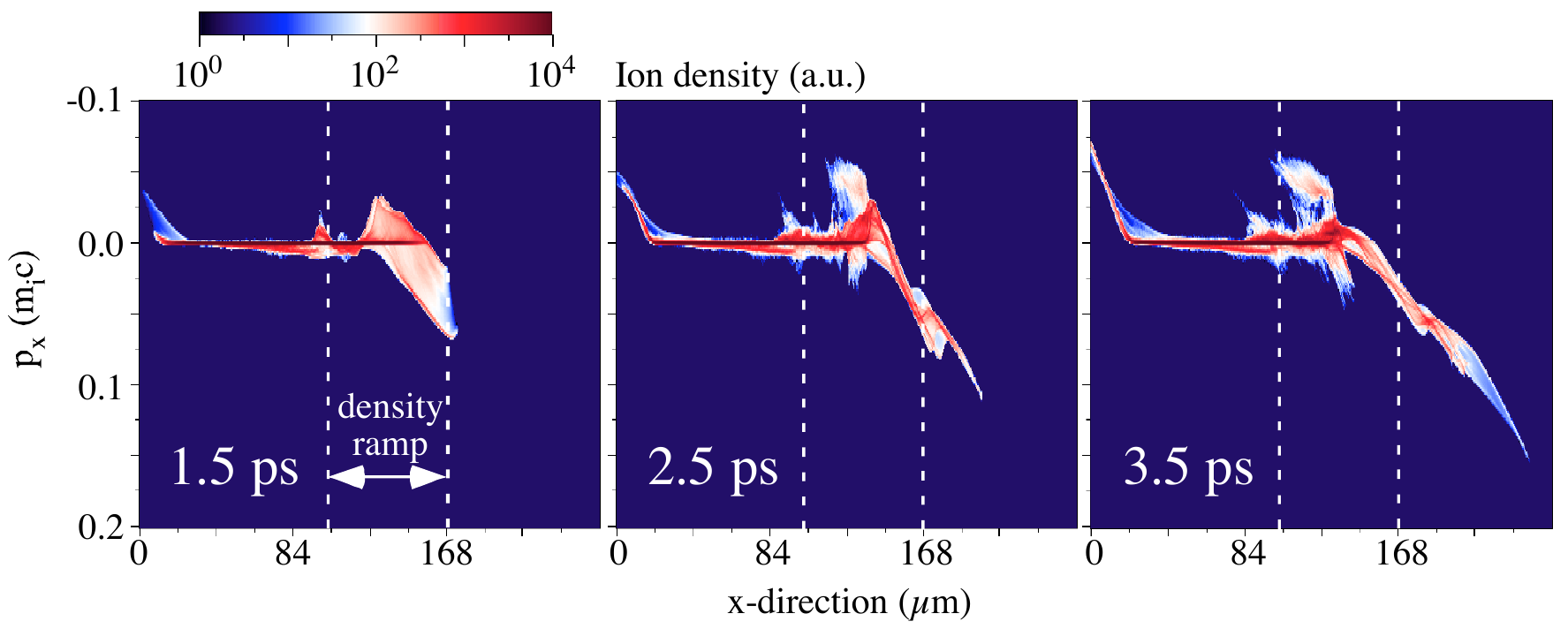}
\caption{$\rm{He}^{2+}$ ion $p_{x}$ against $x$-direction at times of $1.5 \; \rm{ps}$, $2.5 \; \rm{ps}$ and $3.5 \; \rm{ps}$. The initial boundaries of the rear density ramp are shown by the dashed lines.}
\label{figure_p1x1}
\end{center}
\end{figure}

It has previously been suggested that magnetic field effects will be important in similar regimes \cite{Esirkepov_JETP_1999, Bulanov_JETP_2000, Sentoku_PRE_2000, Bulanov_PRL_comment_2007}.
The electron current generates a magnetic field around it and, as it moves into the vacuum, the magnetic field follows.
Magnetic flux is conserved as the plasma expands into the vacuum and this can be seen in fig.\ \ref{osiris_accn_figure} (c).
This time-varying magnetic field contribution to the longitudinal electric field, $E_{\dot{B}}$, can be calculated by studying the temporal evolution of the vector potential due to the quasistatic magnetic field alone using $\nabla \times \mathbf{B} = - \nabla^{2} \mathbf{A}$.
The induced electric field can then be calculated from $\mathbf{E_{\dot{B}}} = - \frac{\partial \mathbf{A}}{\partial t}$ and this contribution at a time of $1 \; \rm{ps}$ into the simulation is shown in fig.\ \ref{osiris_accn_figure} (d) (iii).
The peak value of $E_{\dot{B}}$ at this time is $\approx 0.2 \; \rm{TV/m}$ and this makes up the discrepancy in the longitudinal electric field seen in the simulation, $E_{sim}$.
Furthermore, the contribution $E_{\rho}$ is larger than $E_{\dot{B}}$ at all times in the simulation.

After $1.5 \; \rm{ps}$, a strong radial ($y$-direction) electric field is also observed at the exit of the channel, as can be seen in fig.\ \ref{osiris_accn_figure} (e).
This radial electric field acts to focus ions.
The effect of the focusing force on the ions can be seen in fig.\ \ref{osiris_accn_figure} (f), where the expanding ion front contains structure in the central region.
By analysis of the charge density, it is possible to determine that the radial focusing electric field forms because of charge separation.
The quasistatic magnetic field pinches the electrons, which therefore produces the collimating electric field for the ions.

The ion momentum, $p_{x}$ against $p_{y}$ is shown in fig.\ \ref{osiris_accn_figure} (g) and clearly shows the beam of ions accelerated in the $x$-direction and the radially accelerated ion populations.
The transverse ion momentum, $p_{y}$, is due to the radial ponderomotive Coulomb explosion and begins early in the simulation.
Only once the electrons leave the target and the back surface sheath field is set up, the beam of ions is seen in the longitudinal ($x$) direction.
The $\rm{He}^{2+}$ ion beam has a maximum energy of about $45 \; \rm{MeV}$ and a divergence of $7^{\circ}$ half angle at $25 \; \rm{MeV}$ at the end of the simulation.

The ion acceleration region is illustrated in fig.\ \ref{figure_p1x1}, which shows the longitudinal momentum, $p_{x}$, against the $x$-direction for times into the simulation of $1.5 \; \rm{ps}$, $2.5 \; \rm{ps}$ and $3.5 \; \rm{ps}$.
The dashed lines indicate the initial rear density ramp boundary.
From this it can be seen that all of the high energy ions accelerated in the forward direction originate from the back of the density ramp near the plasma vacuum interface.
It is interesting to note that the highest energy ions are accelerated from a significant depth into the density ramp.
This suggests that the Debye length in this region is $\sim 10 \; \mu \rm{m}$.

The azimuthal magnetic field will exert a magnetic pressure, $P_{B} = B^{2} / 2 \mu_{0}$, on the electrons, preventing the return current, which will allow the electric field to persist for longer.
Also, this will act to enhance the accelerating field for the ions by preventing cold electrons from entering the electric sheath region and allowing the ions to feel the accelerating fields for longer \cite{Esirkepov_JETP_1999, Bulanov_JETP_2000, Sentoku_PRE_2000, Bulanov_PRL_comment_2007}.
Holding back the background electrons will keep the electron temperature in the sheath region high, and the electron density low and thereby extending the Debye length in the region.
It is difficult to quantify the overall influence of the magnetic pressure on the ion acceleration process from the simulations.
But it can be deduced that the magnetic field presence is not a requirement for ion acceleration, since 1D simulations show similar ion acceleration for a purely space charge generated electric field.

\subsection{Effect of plasma density}
\begin{figure}
\begin{center}
\includegraphics[width=7cm]{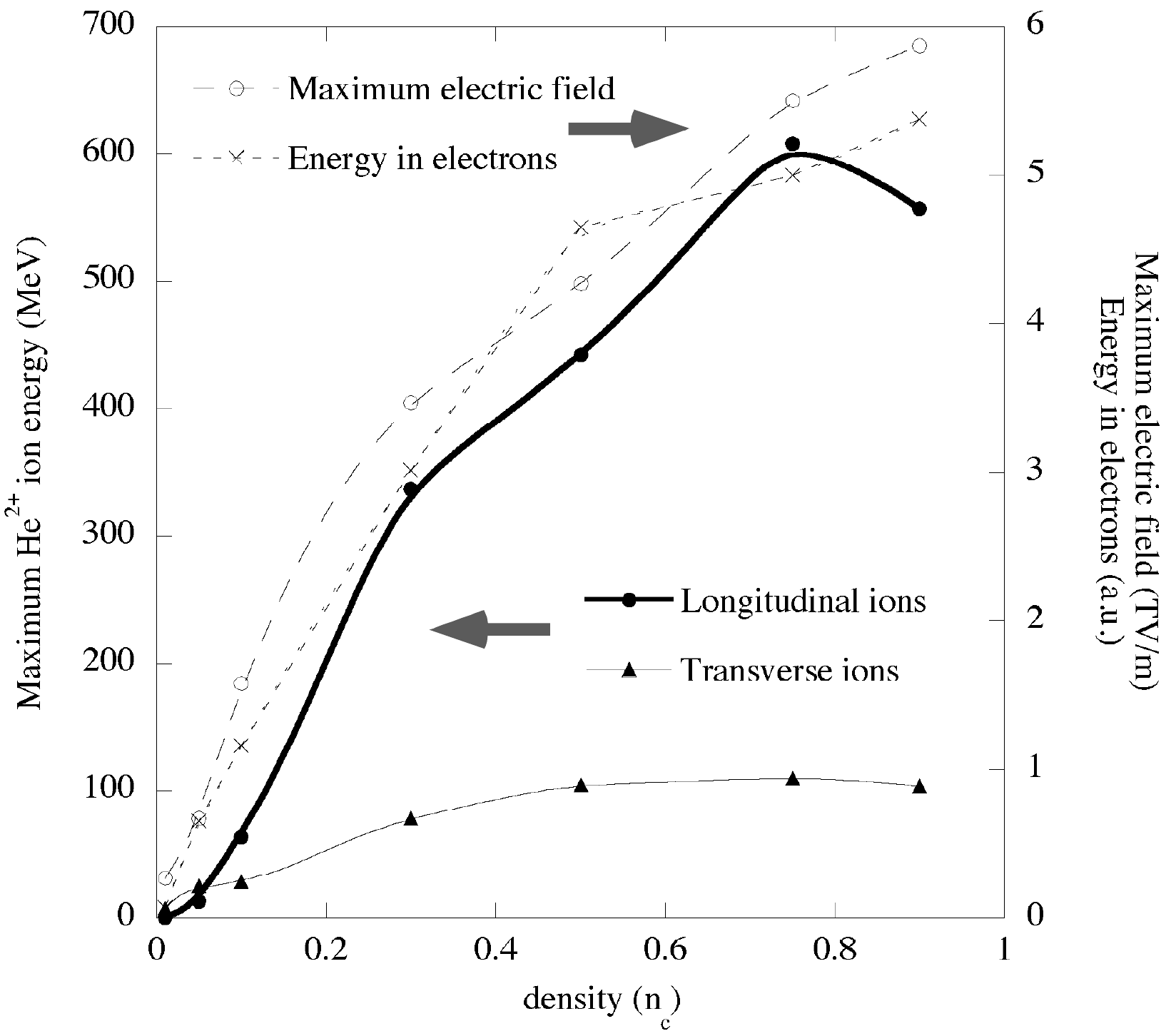}
\caption{Maximum $\rm{He}^{2+}$ ion energies in both the longitudinal and transverse directions at $1.9 \; \rm{ps}$ into the simulations, the maximum electric field (not at the same times in the simulations) and the total longitudinal energy in the electrons ($> 10 \; \rm{MeV}$) for each of the densities. The lines are a smoothed curve fit to the data.}
\label{figure_osiris_energy}
\end{center}
\end{figure}

A number of simulations were performed to investigate the influence of the plasma density on the ion acceleration.
The density scan reveals the same dependance as that observed experimentally.
At the lowest density ($0.01 n_{c}$), the ion acceleration is almost purely in the transverse direction.
However at $0.05 n_{c}$ the longitudinal acceleration is already more effective than the radial acceleration.
Fig.\ \ref{figure_osiris_energy} shows the maximum $\rm{He}^{2+}$ ion energies reached in the transverse and longitudinal direction at $1.9 \; \rm{ps}$ into the simulation as a function of simulated plasma density.
In the transverse direction, the ions are accelerated beyond the ponderomotive potential, $U_{p} = m_{e} c ( \sqrt{1 + \left< a^{2} \right>} - 1) \approx 5 \; \rm{MeV}$ for peak $a_{0} = 15$, for all densities simulated.
This can be attributed to self-focusing of the laser (enhancing $a_{0}$) \cite{Krushelnick_PRL_1999} and shock acceleration, as previously reported by Wei et al.\ \cite{Wei_PRL_2004}.

The longitudinal acceleration increases with density until it peaks at $0.75 n_{c}$, before a slight reduction as the density is increased to $0.9 n_{c}$.
As the density increases, there are more electrons available for acceleration and therefore density of exiting electrons on the rear surface increases.
In figure \ref{figure_osiris_energy} this is illustrated by the maximum electric field seen in the simulation and the total longitudinal energy in the fast ($> 10 \; \rm{MeV}$) electrons, which can be seen to increase with density.
This leads to an increase in the electric field and is able to accelerate ions to higher energies.

\subsection{Effect of plasma ramp length}
\begin{figure}
\begin{center}
\includegraphics[width=9cm]{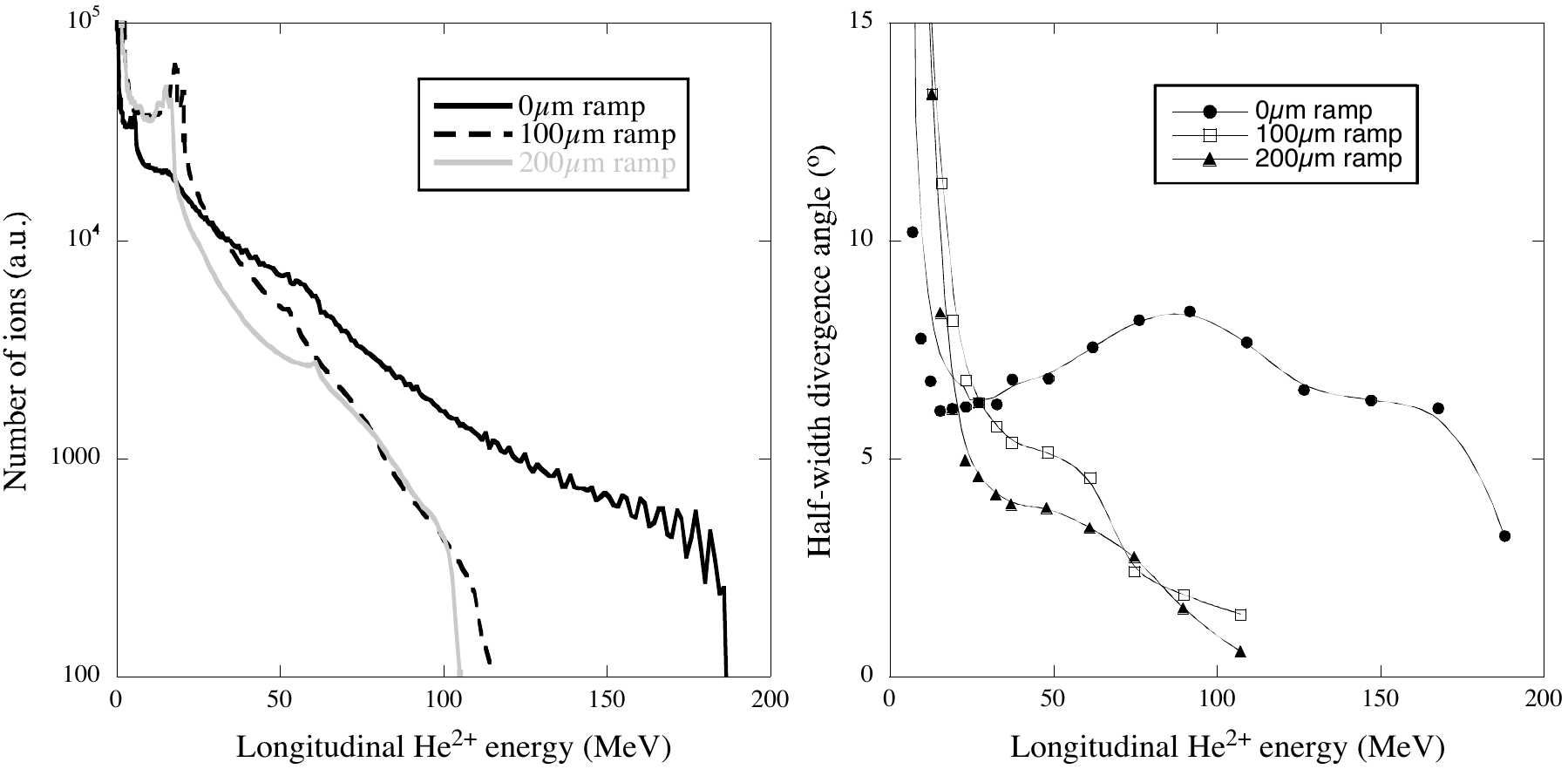}
\caption{The simulated $\rm{He}^{2+}$ ion spectra (left) and the half-width divergence angles (right) as a function of energy for the different ramp lengths at a time when the ion front has expanded by $100 \; \mu \rm{m}$.}
\label{figure_ramp_plot}
\end{center}
\end{figure}

To investigate the effect of the plasma ramp length, simulations with rear density ramp lengths of $0 \; \mu \rm{m}$, $100 \; \mu \rm{m}$ and $200 \; \mu \rm{m}$ were performed.
Due to the different positions of the plasma vacuum interface for the different ramp lengths, the ions start to accelerate at different times into the simulation.
The simulations are compared at the time when the ion front has moved $100 \; \mu \rm{m}$ from the initial rear plasma vacuum interface.
For the $0 \; \mu \rm{m}$ ramp this is at $3.2 \; \rm{ps}$, for the $100 \; \mu \rm{m}$ ramp this is at $3.9 \; \rm{ps}$ and for the $200 \; \mu \rm{m}$ ramp this is at $4.4 \; \rm{ps}$.
Fig.\ \ref{figure_ramp_plot} shows the longitudinal ion spectra (left) and half-width angular divergence (right) of the ion beam as a function of energy at a time when the ion front has expanded by $100 \; \mu \rm{m}$.
The maximum electric sheath fields seen in the simulations are; $2.4 \; \rm{TVm}^{-1}$ at a time of $1.7 \; \rm{ps}$ for the $0 \; \mu \rm{m}$ ramp, $1.3 \; \rm{TVm}^{-1}$ at a time of $2.2 \; \rm{ps}$ for the $100 \; \mu \rm{m}$ ramp and $1.2 \; \rm{TVm}^{-1}$ at a time of $2.4 \; \rm{ps}$ for the $200 \; \mu \rm{m}$ ramp.
As expected, the shortest density ramp allows the acceleration of the highest energy ions.
However, it does not produce the most collimated ion beam, with the $100 \; \mu \rm{m}$ and $200 \; \mu \rm{m}$ ramp lengths showing a significantly better collimation in the high energy tail than the $0 \; \mu \rm{m}$ ramp.

\begin{figure}
\begin{center}
\includegraphics[width=6cm]{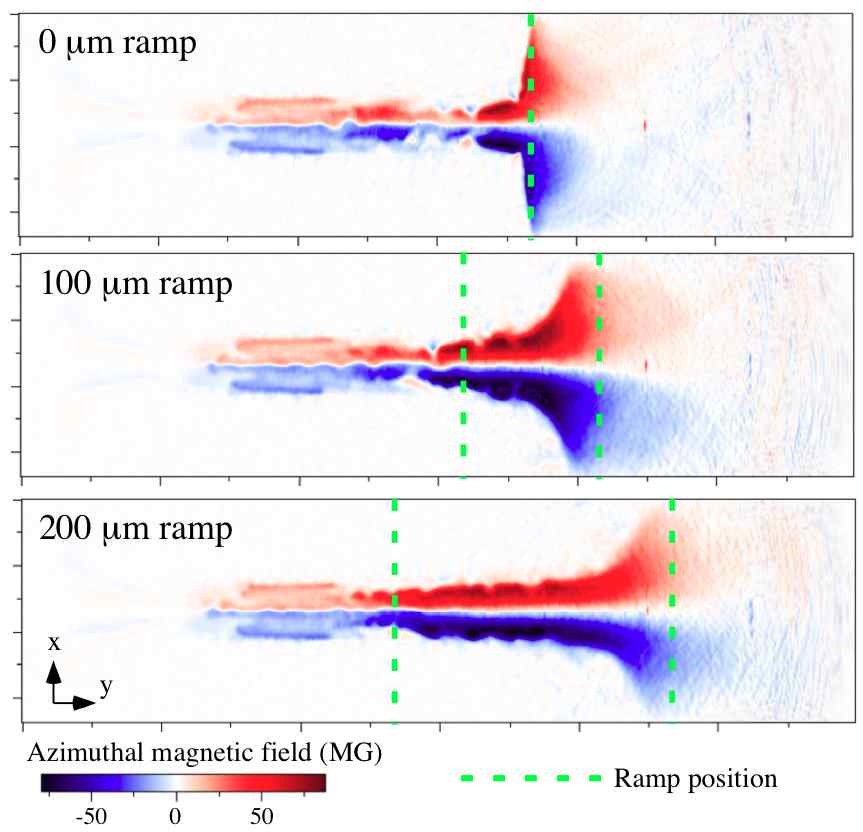}
\caption{The azimuthal magnetic field at a time of $2.2 \; \rm{ps}$ into the simulations for each of the ramp lengths. The initial rear density ramp positions are shown for each simulation. The simulation box is $598 \times 168 \; \mu \rm{m}$.}
\label{figure_ramp_pictures}
\end{center}
\end{figure}

To investigate the improved collimation for longer ramp lengths, we need to consider the azimuthal magnetic field.
Fig.\ \ref{figure_ramp_pictures} shows the azimuthal magnetic fields at a time of $2.2 \; \rm{ps}$ after the start of the simulation for each ramp length.
The longer ramps sustain a longer and stronger azimuthal magnetic field, which will increase the pinching of electrons and provide a stronger focusing force for the ions.
The maximum magnetic fields seen in each of the simulations are $88 \; \rm{MG}$ for the $0 \; \mu \rm{m}$ ramp at $2.1 \; \rm{ps}$, $102 \; \rm{MG}$ for the $100 \; \mu \rm{m}$ ramp at $2.0 \; \rm{ps}$ and $85\; \rm{MG}$ for the $200 \; \mu \rm{m}$ ramp at $2.3 \; \rm{ps}$.
Even though the maximum field is slightly lower for the $200 \; \mu \rm{m}$ ramp, the energy stored in the magnetic field is greater as it is sustained over a longer length.

\section{Discussion}

\subsection{Comparison with sheath acceleration models}
Consider a simple plasma expansion into a vacuum model of ion acceleration by an electrostatic sheath \cite{Mora_PRL_2003, Gitomer_PoF_1986}, which gives an expression for the maximum ion energy, $\epsilon_{max} \approx 2 Z k_{B} T_{e} \left[ \ln \left( \tau + \sqrt{\tau^{2} + 1} \right) \right]^{2}$, where $T_{e}$ is the fast electron temperature, $\tau = \omega_{pi} t / \sqrt{2 e_{E}}$, $t$ is the time that the electrons are hot and $e_{E}$ denotes the numerical constant $2.71828 \ldots$.
The model implies that the maximum ion acceleration achievable would improve with increasing the fast electron density ($\omega_{pi} \propto \sqrt{n_{e}}$), temperature, $T_{e}$ or acceleration time, $t$.
Taking a plasma density of $4 \times 10^{19} \; \rm{cm}^{-3}$ with a measured electron temperature of $7.4 \; \rm{MeV}$ and a laser pulse length of $1 \; \rm{ps}$, the model gives a maximum $\rm{He}^{2+}$ energy of $117 \; \rm{MeV}$.
However, the model assumes a sharp plasma vacuum boundary, which was not the case for the presented experiment and a density ramp is expected to reduce the maximum ion energy.

\subsection{Laser parameter effects}
As the laser energy increases, it has been observed experimentally that the high energy electron spectra becomes hotter and contains more electrons \cite{Nagel_RAL_2006}.
The plasma expansion model therefore suggests that increasing the laser energy would be favorable to the ion acceleration.
However, as the intensity is increased, the laser can break up into a number of filaments \cite{Najmudin_PoP_2003}.
Due to the filamentation, there is a lower peak intensity but electrons can still be accelerated inside the individual plasma channels.
This leads to the electron beam exiting the plasma in a number of lower temperature beams from a larger area.
Due to computational constraints, the size of the simulation box used for the investigations performed here are not large enough to simulate the entire plasma length, which is presumably why filamentation of the laser beam did not occur.
However, in similar parameter, but longer plasma, filaments are often observed.

Although the effect of varying the laser pulse duration, $\tau_{L}$, has not been studied here, assuming that the laser pulse length is directly related to the acceleration time, the plasma expansion model suggests that a longer pulse length would lead to higher energies.
However, lengthening the laser pulse would not necessarily improve the acceleration since for fixed laser energy the intensity would decrease and give less efficient electron acceleration.
Studies seem to imply that $T_{e} \propto a_{0} \propto \sqrt{\tau_{L}}$ \cite{Nagel_RAL_2006}, whereas lengthening the pulse may lead to an acceleration time increase $\propto t$.
Therefore, there is likely to be an optimum pulse length for a particular laser energy, to maximize the ion energies and this has yet to be investigated.

\subsection{Plasma parameter effects}
Both the experiment and simulations show that the longitudinal ion acceleration improves with increasing ion density (to a certain point).
It is expected that more electrons would be accelerated from a higher density plasma, leading to stronger accelerating sheath fields and this is shown in figure \ref{figure_osiris_energy} as the total number of electrons accelerated.
On the other hand, the electron temperatures have been observed experimentally to drop at higher plasma density \cite{Nagel_RAL_2006}, which is possibly due to filamentation of the laser pulse.
It is probable that the electron acceleration is overestimated in these 2D simulations as effects such as self-focusing are reduced and that is expected to lead to less filamentation.
A reduction in the rate of the acceleration improvement is seen in the maximum ion energy starting at $0.3 n_{c}$, which could be related to the laser propagation.
The group velocity of the laser, $v_{g} = c \sqrt{1 - (n_{e}/ \left< \gamma \right> n_{c})}$, is slower in higher density plasma.
This leads to the electron current outrunning the laser pulse.
Not only does this reduce the energy the electrons can gain, but also the electrons arrive at the rear surface before the laser fields.
Without the laser fields, the assistance of the ponderomotive force in removing the background electrons is not present and the sheath field would be reduced.
This is a possible explanation for the reduction in maximum ion energy as the density approaches the critical density.

For solid density targets, increased rear-side density scale-length, $L$, has been shown to dramatically reduce the acceleration, since the maximum electric field is given by $E_{max} \approx T_{e} / (eL)$ \cite{Wilks_PoP_2001, Kaluza_PRL_2004}.
The Debye length, $\lambda_{D}$, for solid target plasma densities is  on the order of nanometers.
The scale-length due to the initial gas density ramp in both the experiment ($\sim 250 \; \mu \rm{m}$) and the simulations ($71 \; \mu \rm{m}$) is larger than the Debye length, which is estimated to be on the order of micrometers.
Therefore, by the previous reasoning, the relatively long scale-length for the underdense plasma may be expected to reduce the maximum electric field by greater than an order of magnitude.
On the other hand, the magnetic pressure on the background electrons, which are impeded from entering the sheath field region, will have the effect of keeping the Debye length in this region relatively long.
Another interesting effect associated with the rear plasma ramp length is the improved collimation of the ion beam with increasing ramp length as seen in the simulations, which is again a consequence of the magnetic field generation.

\section{Summary}

A collimated beam of ions was measured from an experiment performed using a high-intensity laser interacting with a helium gas jet target.
Simulations have shown that the ion acceleration mechanism was target normal sheath acceleration with contributions to the electric field from both charge separation and the time-varying magnetic field.
The influence of the time-varying magnetic field on the interaction and ion acceleration is expected to be of greater importance for ultra-short ($\tau_{L} \sim \rm{fs}$) petawatt class laser interactions.
The maximum ion energy is expected to increase with plasma density due to the larger maximum electric field generated by the greater electron current.
Indeed, the simulations suggest that very high energy ions could be accelerated if sub-critical, but close to critical densities could be accessed.

\section*{Acknowledgment}

The authors acknowledge the staff of the Central Laser Facility (RAL) for technical assistance.
We gratefully acknowledge the Osiris consortium (UCLA/IST/USC) for the use of Osiris.

$^\dag$ Present address: Fusion Science Center and Laboratory for Laser Energetics, University of Rochester, NY, USA.
$^\ddag$ Present address: Institute for Optics and Quantum Electronics, Jena, Germany.
$^{\ast}$ Present address: Center for Ultrafast Optical Science, University of Michigan, Ann Arbor MI, USA.

\end{document}